# Hybrid Optical and Electrical Network Flows Scheduling in Cloud Data Centres

Ibrahim Kabiru Musa[1] and Stuart Walker[2]

Department of Computer Science and Electronic Engineering
University of Essex, Colchester, UK, CO4 3SQ
email:[1]ikmusa@essex.ac.uk,[2]stuart@essex.ac.uk

## ABSTRACT

*Hybrid intra-data centre networks, with optical and electrical capabilities, are attracting research interest in recent years. This is attributed to the emergence of new bandwidth greedy applications and novel computing paradigms. A key decision to make in networks of this type is the selection and placement of suitable flows for switching in circuit network. Here, we propose an efficient strategy for flow selection and placement suitable for hybrid Intra-cloud data centre networks. We further present techniques for investigating bottlenecks in a packet networks and for the selection of flows to switch in circuit network. The bottleneck technique is verified on a Software Defined Network (SDN) testbed. We also implemented the techniques presented here in a scalable simulation experiment to investigate the impact of flow selection on network performance. Results obtained from scalable simulation experiment indicate a considerable improvement on average throughput, lower configuration delay, and stability of offloaded flows..*

## KEYWORDS

*Cloud computing, Optical Network, Hybrid Network, Openflow, Laplace Graph*

## 1. INTRODUCTION

Emerging data and computer intensive applications, driven by new computing paradigms, are exerting pressure on traditional data centres. One example of these models is cloud computing which offers a new economically viable approach of delivering network and IT resources as services. Cloud computing allows the use of resources on pay-as-you-go basis eliminating upfront investments on capital and enabling users to scale up and down to meet unexpected events [1]. Cloud service scalability must be achieved while protecting Service Level Agreement (SLA) such as consistent latency, high performance, and flexibility. Bandwidth over-provisioning and over subscription, used as techniques to meet these challenges, work only with prior understanding of future traffic characteristics; a somewhat idealised situation [1]. However, cloud computing operates on dynamic basis where traffic is usually unknown in advance and difficult to determine. A possible approach is to build a reconfigurable hybrid network with dynamic control and management planes. Hybrid network makes sense considering that Electrical Packet Network (EPN) is suitable for bursty flows due to its statistical multiplexing benefits. However, packet networks may not solve the requirements of bandwidth hungry applications and do not have good scalability. Optical Circuit Networks (OCNs) are suitable for high bandwidth, long duration, and stable and pairwise traffic [2]. Examples of such traffic includes massive data transfer [1] [3], 3D on-line gaming, and high performance scientific applications [1] [4]. Providing optical paths for low duration and delay sensitive flows may not be realistic due to the longer configuration delay - in the range of hundreds of milliseconds. Typical examples of such traffic are Enterprise Content Management [5] , web search, business transactions, and alert systems.

Network scalability can be achieved by offering low capacity EPN initially which dynamically scale up to higher capacity OCN. This creates the experience of seemingly infinite cloud resources. This way both coarse and fine grain traffic sizes can be accommodated. POD [6] and modular [2] based hybrid data centres offer such designs. Hybrid networks require effective strategies to meet the constraints of both OCN and EPN. In OCN, for example, the maximum number of circuits to be created at each node is bounded by half the number of elements to connect and the setup delays. EPN on the other hand, require high port count and large memory buffers for queuing, as is evident to switching delays. Another challenge is the introduction of additional layer of network complexity by virtual machines (VM) running inside hosts. As a result, virtual machine sprawl means that multiple VMs can run in one host creating the possibility of bottlenecks. Research into balanced traffic flow distribution across EPN and OCN to utilize the capabilities of both network types is now attracting interest. Critical decisions to make are the determination of traffic conditions to necessitate EPN flow offload, for example.

In this paper we propose a mechanism for dynamic traffic change detection and flow scheduling suitable for EPN traffic offload into OCN in an intra-Cloud Data centre Network (CDC). We argue that selecting a flow suitable for placement into OCN based on traffic contributions and local host traffic impacts has significant impact on performance. We focused specifically on CDC due to the current wide deployment of cloud based applications and services. The rest of the paper is organized in the following manner. Section 2 gives an overview of cloud data centre, section 3 review related works, section 4 describe our proposed flow scheduling strategy, section 5 presents the results of our experiment, and section 6 concludes the paper.

## 2. OVERVIEW ON CLOUD DATA CENTRE

Cloud computing offers a type of parallel and distributed system where collection of inter-connected and virtualized computers are dynamically provisioned and presented as one or more unified computing resource(s) based on Service Level Agreements (SLA) [7]. This is established through negotiation between the service provider and user [1]. In this model, distributed resources are orchestrated into a large centralized pool enabling ubiquitous, convenient, on demand multiple network access to a shared pool of configurable computing resources (e.g. networks, servers, storage, applications, and services). The resources should be rapidly provisioned and released with minimal management effort or service provider interaction [4]. Traditionally, large scale dedicated infrastructure is organized as a Data Centre (DC) to support these cloud services. However, data centres for true multi-tenant cloud computing need to provide mechanisms for enhancing application performance by ensuring rapid deployment of networking services and supporting high scalability, flexibility, high availability, and rapid resilience [1]. This emerging type of DC is commonly referred to as Cloud DC (CDC).

 A major challenge of CDCs is the need for efficient resources utilization and virtualization. Virtualization technologies create additional layers of complexity at the host level as a result of virtual networking. Another challenge is the service-oriented resources view in cloud computing aim at scale out to reduce under-utilization and maximize revenue returns. There is also the need to maintain application performance for each tenant in the cloud as agreed in SLA. CDC must also ensure rapid deployment of networking services. Furthermore, CDC must provide high scalability and flexibility to support applications' need to access services on demand with minimum intervention of service provider. Most importantly, functional requirements of emerging applications such as mobile and interactive, parallel processing, visual gaming, streaming, and analytical analysis all must be taken into consideration in future CDCs.

## 2.1. Network requirement for cloud computing

Networks constitute an indispensable resource in cloud computing. First, cloud services are accessed over a network (internet, private networks etc.) linking end users and cloud services at a provider's edge and ensuring ubiquitous access. Secondly, cloud services are created in data centres using local (or federated) infrastructure comprising large computing and storage resources connected by a network. Currently, there is upsurge in the bandwidth requirements of applications deployed in data centres. This is attributed to penetration of the internet and emergence of bandwidth greedy applications. Video and voice are at the forefront of this change and require gigabytes of network capacity in large user number environment. Optical networking is poised to support these requirements requiring less operational cost to connect disparate resources using wide variety of interfaces, offering huge capacity, and efficient physical properties. Efficiency in space and power consumption, lower management [8] complexity, lower maintenance costs are core features of optical networks. All these features contributed to the interest in optical networking for cloud computing.

However, for cloud computing to maintain the notion of accommodating all service types, there is a need to support different applications each with specific requirement for bandwidth granularity. All requests are serviced with minimal provider or user intervention with high elasticity. Request elasticity, on-demand provisioning, and high scalability are combined to achieve satisfactory level of latency and response time for each cloud tenant..

## 2.2. Various network types in cloud data centres

Cloud data centres consist of three important networks. Access network between user and cloud resources, Intra Cloud Data Centre Networks (DCNs) connecting infrastructure within a data center and Inter Cloud Data Centre Networks (DCIs) connecting multiple clouds or infrastructures residing across large geographical distance. Existing DCN infrastructure are usually large geographically, covering 10 km or more and hosting a wide variety of applications of various sizes with a requirement for high scalability. Currently, OCNs are widely deployed in DCI and EPN in DCN. EPNs dominates DCNs due to their statistical multiplexing capabilities, support for per-flow switching, and capability to handle short duration flows and bursty traffics [8] [2].

OCN dominates DCI due to its long reach, rate agnostic switching, suitability for aggregate flows with less space requirements, less error rate, and high energy efficiency [8]. It is thus inefficient to apply OCN at the granularity level of end hosts that transmit less than 10 GB/s and carry out bursty communication [2]. Recent research studies [8] suggest the combination of EPNs and OCNs to accommodate all requests from small to extremely large scales in a controlled manner. Cost, business agility, efficiency, and objective latency for various traffic types are also achieved as a result of this combination. CDC host various flows of different size and pattern with long lived flows greater than 100ms in intra data centres [9]. Hence, majority of the flows are suitable for EPNs. Achieving a balance of distribution is therefore critical to successful deployment of CDC hybrid network.

## 2.3. State of the art in optical technology

High capacity optical data transmission emerged in early 70's and rapidly gained acceptance with standardizations by The Optical Network Technology Consortium (ONTC), Multi-wavelength

Optical Networking (MONET), and The European Multi-wavelength Transport Network (MTWN). Short and long haul gradually enhanced to provide Fibre To The Desk and Home (FTTH and FTTD), as well as and Gigabit Ethernet [10]. Increase capacity was achieved by multiplexing wavelengths in Dense and Coarse WDM, and DWDM-RAM. Dynamic characteristics were achieved through multi-degree ROADM (Reconfigurable Optical Add and Drop Multiplexer) and Optical Cross Connects (OXC). These technologies addresses DWDM shortcomings by allowing networks to be designed with an optical dynamic core, offering network flexibility, support for complex mesh topologies, and reconfigurable light paths. As more components are considered for deployment, control and signalling mechanisms become important compliments to allow network awareness, provisioning, and multi-layer technologies. Large scale research into Generalized Multi-Protocol Label switching (GMPLS) as a viable control mechanism is widely described in the literature [11] [12]. Research into Optical Burst Switching (OBS), which combines the transparency of Optical Circuit Switching (OCS) with the statistical multiplexing gain of Optical Packet Switching (OPS), presents new and enhanced techniques for signalling and wavelength assignment [12].

As next wave of technologies impact on everyday scientific and social trends, optical network technology is expected to dominate communication systems. Technologies such as cloud computing, social networking, high visual gaming applications, and high resolution video streaming are expected to exert more pressure on optical network technologies.

## 2.4. Typical DCN traffic scenario

Cloud service tasks are submitted in interdependent form. This is usually a cyclic relationship where overall tasks are achieved by completing multiple subtasks. This collection of task is generally viewed as Bag of Task (BOT). Each subtask is performed by scheduled virtual or physical DCN resources as indicated in figure 1. Each resource is hosted in a Physical Host (PH). Physical hosts are connected by a finite Physical Link (PL). In figure 1, resource $R_1$ requires resources $R_2$ and $R_3$ to complete task $T_1$. $R_2$ in turn requires a resource $R_3$ to complete task 2. And finally resource $R_1$ requires $R_2$ to complete task 3. The output from the processes are henceforth $O_1 = \{R_1, R_2, R_3\}$, $O_2 = \{R_2, R_3\}$, and finally $O_3 = \{R_1, R_2\}$. For N number of flows we

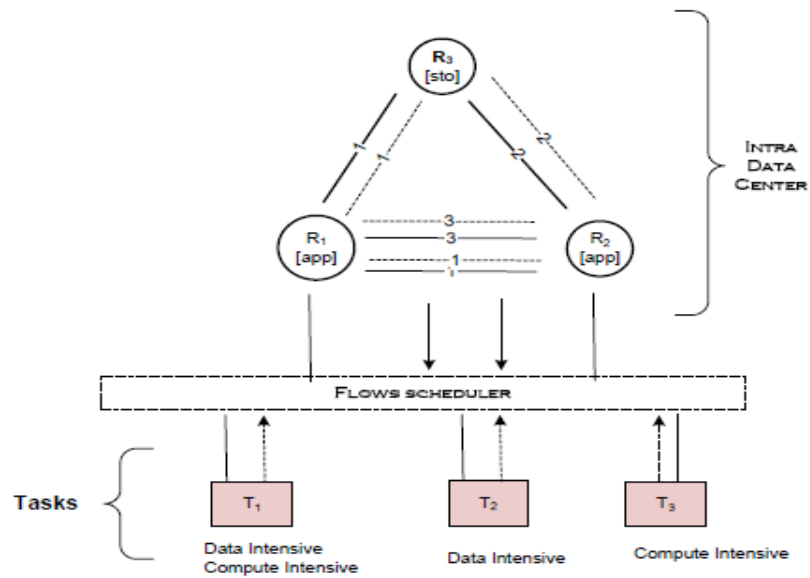

Figure 1. Demonstrating the resource interactions in Intra Cloud Datacenter (DCN)

have the total resource request transmitted as $\sum_{i,j \in R} \sum_{m \in N}[(R_i * X^m_{i,j})]$. Where $X^m_{i,j}$ indicates the existence(or non) of interaction between resources index i and j to complete task m. N can range from 100s to 10s of thousands. An example of such a BOT scenario is found in multimedia cloud [13] where cloud users request video streams from a provider. The request is delivered to an access server ($R_1$ in figure 1). The access server identifies the user device as well as other request parameters. The access server then, downloads video content from a storage server ($R_3$). The content is then encoded to meet specific requirement by another server ($R_2$). The final output sent to the user is the encoded video that satisfies device limitations and Quality of Service (QOS) requirements

Next we present an overview of some related works available in the literature and the major contribution of this work.

## 3. RELATED WORKS

Trends in Cloud Computing debut the emergence of data intensive applications [4] and large diameter data centres. This trend necessitates the combination of Electrical Packet Network (EPN) and Optical Circuit Network (OCN) into a hybrid DCN [2]. This view is becoming an active research topic described in the literature [2] [6]. The approach is to design data centre Network (DCN) with EPN handling short lived and bursty flows leaving the long lived and data-intensive flows to OCN. This way scalability to support applications needs in accessing cloud services on demand can be achieved. In [2] the access layer connecting host is viewed as a POD. Each POD contains both transceiver and optical multiplexer for packet and circuit networks respectively. Traffic flows are assigned using a random time variant scheme where number of hosts in a POD are periodically assigned optical switching network. The study assumed a bijection traffic model where each host receives an amount of traffic equal to the total traffic sent by the host. It was shown in the research that stability of traffic flows have significant impact on the throughput. Hedera [14] proposed a flow scheduling which dynamically estimate loads and then move flows from heavily loaded links to less utilized links. A scheduler ensures that flows are assigned non conflicting paths and optimum bisection achieved. Another important trend pushing this approach is the fall in cost of Optical transceivers [8].

The aforementioned research studies identified interesting phenomenon on cloud services and hybrid network design for CDC. Our work addresses two aspects not fully addressed in the literature - dynamic mechanism to detect network changes in hybrid DCN and optimal flows offload from EPN into OCN. This selection is the first step in building hybrid network suitable for CDC and eventually affects the suitability of flows after offload into OCN. The main contribution of our work is the development of strategy for selecting and placement of suitable flows for switching in optical or electrical domain based on statistical traffic contributions. This is different with the approach where random assignments are used to switch flows in electrical or optical network. Our work proposes the use of software define networking (SDN) to achieve dynamic configuration of communication devices. We were able to integrate optical switching capabilities into a widely used scalable simulation platform, CloudSim [15], and run a simulation to test our proposed scheduling algorithm. The approach for the selection of flows in our work is based on graphical structure of a traffic statistics graph. Statistics considered are the logical and traffic amount dominance of flows.

## 4. FLOW SCHEDULING FOR CLOUD INTRA DATA CENTRE

Our architecture is depicted in Figure 2 and we assume the following:
- Packets network is designed as a simple heavy tail model.

- Wavelengths are regarded as limited and scarce resources.
- No wavelength conversion in circuit networks
- Non negligible time is required to create and release optical route from source to destination
- Basic Openflow protocol install at host access level switches and not on circuit switching devices to allow all optical switching (OOP) behaviour.

We formulate the architecture for dynamic traffic management in hybrid electrical and packet switching network.

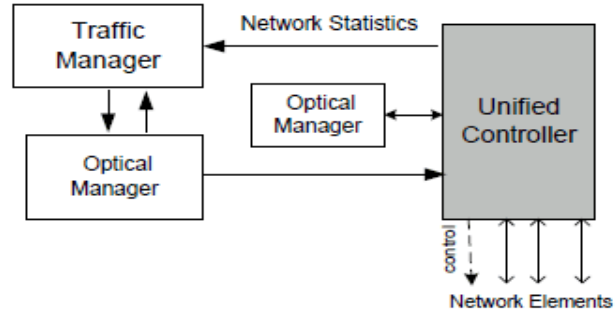

Figure 2. Functional units involved in detection and selection of flows suitable for unload into Optical Network from Packet Network

The functional structure of our proposed architecture is depicted in Figure 2. It comprises a controller, Traffic Manager, and optical manager.

---

**Algorithm 1** Pseudo code For Traffic Change Computation and flow scheduling

**Input:** Traffic statistics S,D,$\omega$, where S is source set, D is destination set and $\omega$ traffic contribution, N: the set of nodes in the network

**Output:** Set of Flows $F_{u,v}^{\lambda}$ identified by source u and destination v assigned a wavelength $\lambda$

1: Set Val {// set the acceptable change in traffic graph}
2: Given Non-empty set T= $\{S, D\}$
3: Build graph $g_{u,v}$ and $g'_{u,v}$ from T at $t$ and $t'$ respectively.
4: **for** u,v $\in$ N **do**
5:   compute graph distance $\Delta(g, g')$
6:   **Normalize** $\Delta(g, g')$ over entire network
7: **end for**
8: **if** $max(\Delta(g, g')) > Val$ **then**
9:   Return FALSE {// no bottleneck exit algorithm}
10: **end if**
    {// compute existence of bottleneck}
11: b = $\frac{\sum C_{i,j}}{max(\sum_{u \in \epsilon} d_u, \sum_{v \in \epsilon} d_v)}$
12: **if** $b >= 1$ **then**
13:   Return FALSE {// no bottleneck exit algorithm}
14: **end if**
15: **for** u,v $\in$ N **do**
16:   {// Compute region of dominance}
17:   $d(u,v) = sqrt(\frac{\sum_{i=1}^{k}(\lambda_i(u,v)-\mu_i(u,v))^2}{min(\omega(u,v),\omega'(u,v))})$
18: **end for**{//rank flows based on weight contribution}
19: $d \leftarrow d_\searrow$ {//rank flows in set d in descending order}
20: **for** $F_{u,v} \in d$ **do**
21:   Select $F_{u,v}$ {choose flows based on weight}
22:   mark $T_{u,v}$ from T as selected
23:   Assign wavelength $F_{u,v}^{\lambda}$
24: **end for**
25: **Return F**

International Journal of Computer Science & Information Technology (IJCSIT) Vol 5, No 2, April 2013The flow chart depicted in Figure 3 describes a single pass for our proposed traffic management. The pseudo-code (Algorithm 1) includes the following steps:

- Network nodes send traffic flow statistics updates to centralized controller.
- A traffic manager then generates a traffic matrix with source, destination, and weight identifying each entry.
- A traffic graph is generated with vertices and edges denoting nodes and traffic contribution.
- Selection of suitable flows for offload from EPN to OCN.

A key component of our proposed architecture is a software defined network controller

## 4.1. Software defined network controller

Software Defined Network (SDN) controller implements basic control protocol logic and maintain network statistics and devices information. For our work we use Openflow Controller [16]. Communication between network nodes and controller is through dedicated channel. For simplicity, we assume such communication does not contribute significantly to traffic statistics. This controller query aggregate flow statistics from switches and the switches respond with statistical flow information including table Id, flow count, flow descriptors, out port, in port, packet count, and total byte count. Using SDN controller in a network enables dynamic reconfiguration of communication devices. In our work we integrate a topology manager to provide services to the controller. The topology managers re-compute topology in event of changes in links and node status. New nodes joining the network are also considered in the graph and this triggers re-computation of the traffic graph. This way the controller is aware of the overall network topology making it easy to manipulate forwarding tables with appropriate actions. Every time a packet arrives at Openflow Enabled Switch (OFS) the switch search the flow action table in the switch. The action flow table list flow characteristics and matching action. If there is a policy relating to the new packet an appropriate action is taken on the packet. Action can be to forward the flow to next switch or drop the packet. If there is no matching action for this flow it is encapsulated and send to the controller. Appropriate action is created for the flow and the action tables of the switches updated. Next time the same flow arrive at a switch the action is then applied. For in-depth discussion on SDN and Openflow, in particular, the reader is referred to [16].

## 4.2. Traffic manager

For our model, we define a flows traffic network as a weighted and non-zero graph, G (V, E), consisting of paths connecting X  V and Y  V. Data travel between vertices V by way of the edges E. X is regarded as the input and Y the output. Traffic statistics from each flow is a turple (u, v,  ). For packet size  , source u, and destination v. The traffic manager treats all traffic from node u to node v as a flow $F_{u, v}$. It maintains traffic statistics as n x n matrix such that.

$$A_G = [a_{u,v}] \begin{cases} a_{u,v} = \omega(u, v) & if\ u \neq v \\ a_{u,v} = 0 & Otherwise \end{cases} \quad (1)$$

with entries $a_{u, v}$ defined by an undirected graph g= {V;E;  }, for u, v  V , such that   : V (g) * V (g)  ℝ$^+$, satisfying  (u, v) > 0,   = $F_{u, v}$ = $F_{v, u}$ . We simplify our model as undirected traffic graph by assuming a bijection rule similar to that proposed in [2]. At any two times t and t$^|$ , the traffic matrix generates graphs g and g$^|$ respectively. We compute traffic threshold value $h_T$ as a function of packet network traffic load and link capacity. $h_T$ is the parameter that guarantee flow feasibility. We say that a network is feasible if there exists a flow $F_i(u, v) : V(g) * V(g^|) \to \mathbb{R}^+$ satisfying $F_i(u, v) \leq C(u, v)$ where $C(u, v)$ denotes the capacity available. Four stages are involved in our proposed dynamic traffic scheduling as indicated in figure 3.

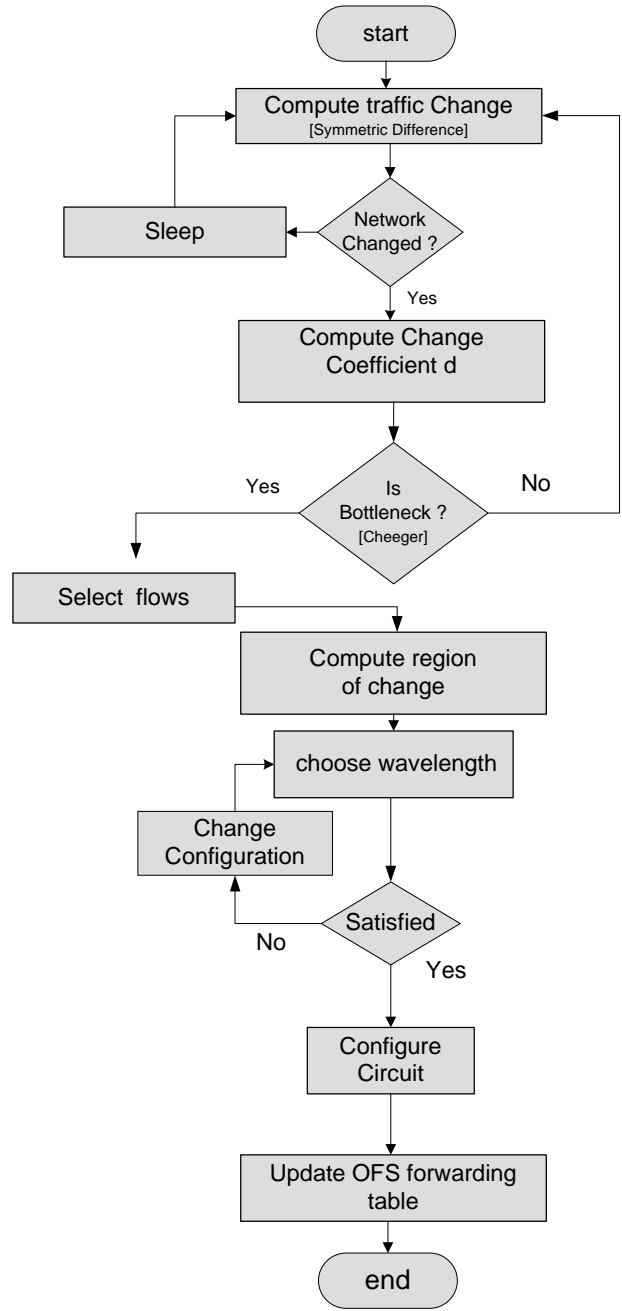

Figure 3. A single pass involved in traffic detection and flow selection in a hybrid optical circuit and electrical packet network

The first step is traffic change detection. This first step avoids the need to compute bottleneck coefficient every time as this is computationally expensive and classified under categories of NP hard problems [17].

International Journal of Computer Science & Information Technology (IJCSIT) Vol 5, No 2, April 2013

### 4.2.1. Stage 1: Traffic change detection

Consider graphs g and $g^|$ with set of edge weights ω and $ω^|$. The relative amount of change over period $\triangledown t = |t-t^||$ between edges is given by $d(g, g^|)$. For any non-zero weights and u,v ∈ V, the traffic change between u and v as the traffic graph evolve from g to $g^|$ is thus:

$$|ω(u,v) - ω^|(u,v)| \qquad (2)$$

Clearly $\max(ω(u,v), ω^|(u,v)) >= |ω(u,v) - ω^|(u,v)|$. And the transformation:

$$d_{(g,g^|)} = \sum_{u,v \in V} \frac{|ω(u,v) - ω^|(u,v)|}{\max(ω(u,v), ω^|(u,v))} \qquad (3)$$

normalizes the traffic change over all the edge sets in V. Once a significant change is detected in the graph, the next stage is started.

### 4.2.2. Stage 2: Estimating the bottleneck coefficient

We adopt Laplace implementation of the Cheeger constant [17] $h_T$ to investigate existence of possible bottleneck in the EPN. This measures the need to offload packets in EPN to OCN. We chose Cheeger constant mainly due to existence of algebraic relationship with proposed Laplacian model adopted in this paper. $h_T$ is estimated to give a measure of the level of bottlenecking in a network flow traffic when modelled as a graph comprising source and destination sets. We particularly use the weighted model of $h_T$. Using the relationship presented in [18] to make provision for traffic edge weight we estimate $h_T$, assuming a traffic bijection model, as $h_T = inf(h)$ where h is defined as:

$$h = \frac{|C(S,D)|}{max_k(\sum_{u \in S} d_u, \sum_{v \in D} d_v)} \qquad (4)$$

Where C(S,D) is capacity set of edges with vertex incident at S. S:set of source nodes, D:set of destination nodes, $d_u$ is the traffic with source incidence at u and not at v, and $d_v$ is the traffic with source incidence at v and not at u. Two cases exists to describe $h_T$ [17]

$h_T < 1$ bottleneck exists
$h_T ≥ 1$ feasible flow exists

Estimating $h_T$ has been shown [19] [17] to be NP-Hard problem since the computation, by definition, can take prohibitive exponential time for small number of vertices. The approach is to obtain an estimate after considerable traffic change is reported in the first stage. The existence of bottleneck triggers the investigation of exact region of change.

### 4.2.3. Stage 3: Identifying dominant traffic region

To detect the contribution by flows to the experienced changes in our graph g and $g^|$, we use the concept of graph Laplace Spectra which measures both the weight and the logical connectivity dominance of a flow. This approach is similar to transport flow measurements introduced in enterprise networks and applied to compute relative changes [18] in real traffic networks such as vehicle traffic network.

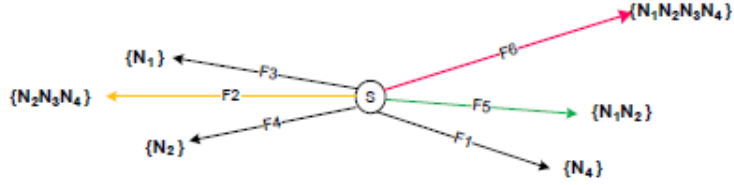

Figure 4. Flow connectivity in intra cloud data center network

flow measurements introduced in enterprise networks and applied to compute relative changes [18] in real traffic networks such as vehicle traffic network. LSG measures both flow contribution and logical flow connectivity. The premise of considering logical flow connectivity is the strong relationship between flow diversity and sustainability over time. Under this a flow with many related flows is expected to last longer making it suitable for OCN switching. Consider the graph in figure 4, flows $F_1$, $F_2$, ..., $F_6$ are all logically connected at source, S. The duration of these flows is a function of data to transport by flow, the maximum transmission size W, and the queuing delay at S. For instance, assume flow $F_6$ contribute large traffic in the network and share network resource $N_1$ with flows $F_3$ and $F_5$. The estimated duration of $F_6$ is given by:

$$T(F_6) = \left(\frac{\Gamma}{W}\right) * \sum \zeta \qquad (5)$$

Where is the service time for each element in the queue. The term on the right of equation 5 contribute significantly to the duration of each flow. To compute specific region of change we first define weighted Laplace matrix as:

$$L(u,v) = \begin{cases} d_v - \omega(u,v) & if\ u = v \\ -\omega(u,v) & if\ u\ and\ v\ are\ adjacent \\ 0 & otherwise \end{cases} \qquad (6)$$

Where $= D^{-1}LD^{-1}$ and $dv = D^{-1}(v;v) = 0$ in the relation. is the Laplace of graph g. The eigenvalues of the Laplace graph has been shown in [17] [20] to be a good and stable indicator of changes in a graph. A Laplace of graph g and spectral index (Ag) = $_0$, $_1$, ..., $_n$ is defined as $_g = D_g - A_g$. Where $D_g$ and $A_g$ are the diagonal matrix and weighted adjacency matrix of g respectively. The Laplace of g is otherwise stated as:

$$(u,v) = \begin{cases} 1 - \dfrac{(u,v)}{d_v} & if\ u = v \\ -\dfrac{(u,v)}{\sqrt{d_v d_u}} & if\ u\ and\ v\ are\ adjacent \\ 0 & otherwise \end{cases} \qquad (7)$$

Let the spectral matrix of g and $g^|$ be given by (Ag) = $_0$, $_1$, ..., $_n$ and (Ag) = $\mu_0$, $\mu_1$, ..., $\mu_n$. n is the number of nodes in the graph at point of computation. And $_0$, $_1$, ..., $_n$ n is the spectrum of the graph. We expect the eigenvalues to change systematically if the graph links and their weights show systematic change. The difference between the spectra provides us with a distance measure between two graphs g and $g^|$. We normalize the differences as in [17] [21] to obtain

$$d(g,g) = \sqrt{\frac{\sum_{i=1}^{k}(\lambda_i - \mu_i)^2}{\min(\sum_i \lambda_i^2, \sum_i \mu_i^2)}} \qquad (8)$$

By considering sub graph g(u, v) such that u, v ∈ E in equation 8 we obtain the Laplace traffic distance as:

$$d(u,v) = \sqrt{\frac{\sum_{i=1}^{k}(\lambda_i(u,v) - \mu_i(u,v))^2}{\min(\lambda(u,v), \mu(u,v))}} \qquad (9)$$

Equation 9 gives specific Laplacian spectral distance for a flow identified by nodes u and v. k=20 have been shown in [20] to be a good value for investigating significant changes in graphs.

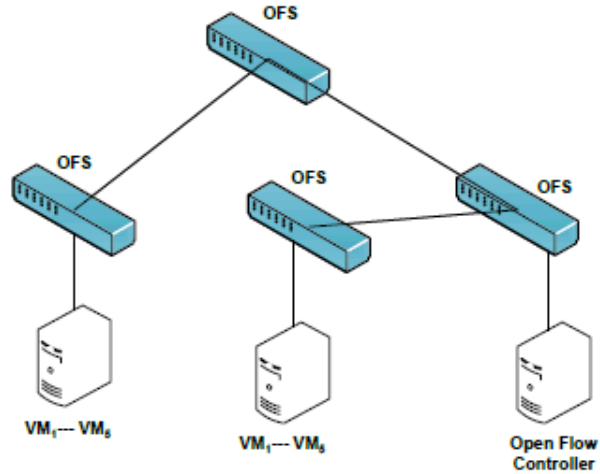

Figure 5. Virtual topology created on Ofelia Openflow-enabled research testbed used to measure various Cheeger constants at varying sample times of the day

### 4.2.4. Stage 4: Ranking identified flows

We model the rank value δ assignment given vector of region of cumulative incremental change computed as:

$$\sum_{i} \lambda_i \qquad (10)$$

For m number of flows contributing traffic λ. We compute rank as:

$$\kappa_{u,v} = \sum \frac{\delta(u,v) - \delta'(u,v)}{\max(\delta(u,v), \delta'(u,v))} \qquad (11)$$

Where λ and λ' are weights contribution at t and t'. Equation 11 rank each candidate flows based on the change in traffic and amount of traffic load contribution. Notice that the rank of a flow increases with normalized value δ(u, v) - δ'(u, v) > 0 and reduces with δ(u, v) - δ'(u, v) < 0. This way only the flows with consistent high traffic load contribution is chosen.

International Journal of Computer Science & Information Technology (IJCSIT) Vol 5, No 2, April 2013

## 5. EXPERIMENTAL RESULTS

We begin the experiment with verification of our bottleneck computation technique described in section 4. For a computer with high computation power, calculating Cheeger constant of a graph with > 20 nodes will take more than 24hrs. We therefore calculate a fast estimate for the Cheeger constant. To establish the computational validity of the Cheeger estimate described in this paper, we created an experimental topology comprising 3 hosts and 4 switches on Ofelia Test bed. Each physical host connect to an Openflow enable layer 2 switch. Each switch connects to a controller which obtains network topology information as well as statistics from all the switches. The test bed is configured to support virtual resources slicing using Flowvisor [22]. This way, virtual topologies are created to connect virtual machines running in the physical hosts. We then ran large UDP stream of sizes of 1000-1500B. Sample of statistics are taken at various time intervals and Cheeger constant with k=20 values computed. Figure 7 shows the pattern of various Cheeger values. Each value is a function of traffic statistics weight and available link capacity. At time stamp 30, in Figure 7, a bottleneck value of 0.86 indicates need to trigger flow offload into OCN. We also investigated the relationship between Cheeger constant and link utilization. 25 sample traffic statistics each over a period of 30s were obtained. Link utilization and Cheeger coefficient were computed and compared.

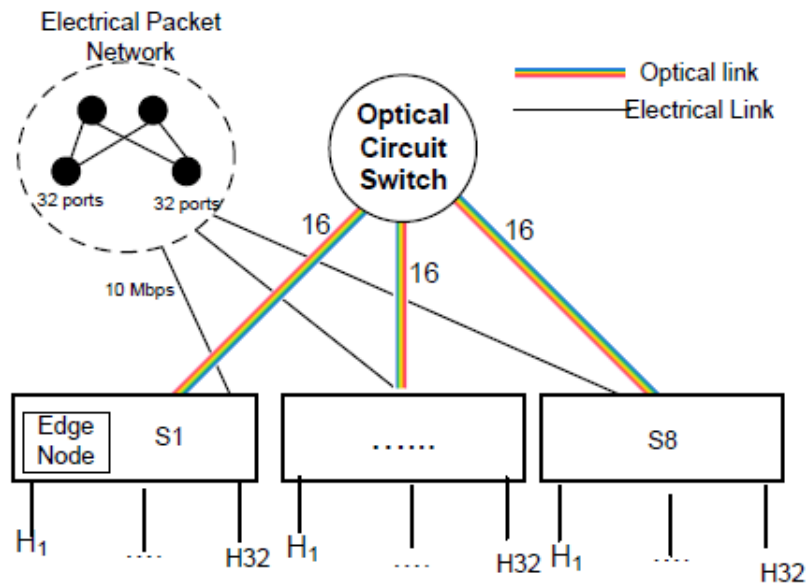

Figure 6. The topology used in the simulation of hybrid optical and electrical intra data centre network with 2 layer networks comprising both Electrical and Circuit switching. Available wavelength is 16 on the optical links

To investigate wide traffic conditions, we stressed the network initially with higher transmission rate and subsequently relaxed the traffic rate. As shown in figure 7 as link utilization reaches a high positive value, the Cheeger constant approaches zero (indicating bottleneck possibility). As the link load reduces, we noticed improvement on the Cheeger values toward higher positive values. These two results show the capability of Cheeger constant to indicate bottleneck in a network.

To investigate the performance of our propose flow scheduling in hybrid EPN and OCN at large scale we implemented a simulation experiment using CloudSim [15] enhanced with optical network modules including optical cross connects, cross connect edge routers, wavelength switching engine, Path computation Elements (PCE), Network Interface Card (NIC), Optical Network Interface Card(ONIC), and channel modules. Optical modules were adopted from Glass project and customized to support the simulation aims. We use the topology shown in figure 6. The topology consists of a packet and optical network. Each access level switch (numbered 1-8) is directly connected to 32 hosts. Each access level switch is connected to both packet and circuit networks. The aggregation packet part of the network contains two electrical switches with 32 ports each. The OCN aggregation networks perform all optical switching. Normalize random traffic arrival is adopted with Poisson for user request and LogNormal distribution for intra traffic modelling. This choice is similar to description of traffic characteristics in data centres investigated in [9]. Each host sends traffic to a random destination using a LogNormal model with shape parameter 0 and scaling between 1-5. Such a model is particularly useful in modelling applications communication pattern in CDC since they are flexible enough to model a variety of data sets. Normalize random traffic arrival is adopted with Poisson model to simulate user request. Where traffic similarity is required, an initial traffic is generated and stored in a file which is then subsequently used in the duration of the experiment.

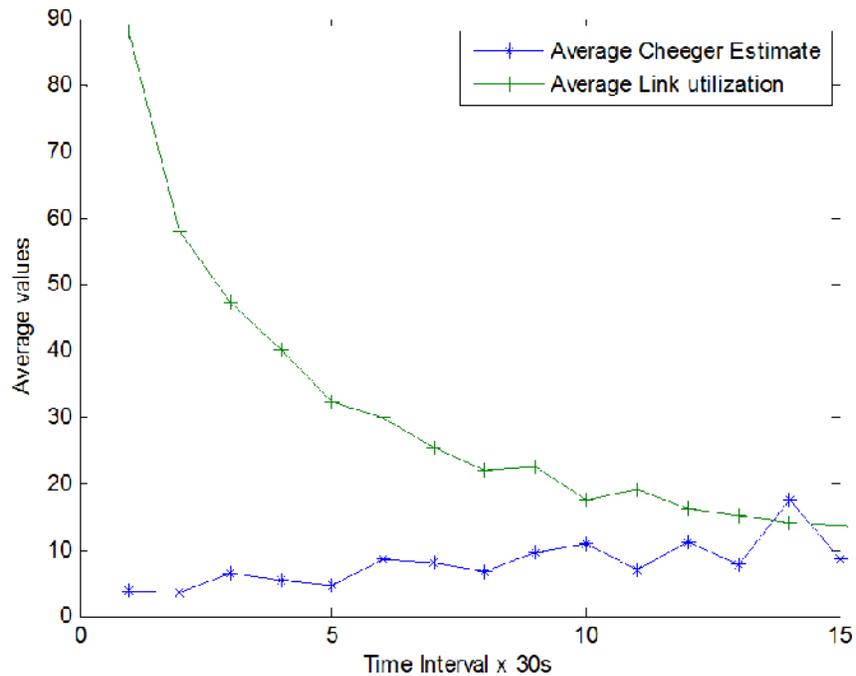

Figure 7. Figure comparing Cheeger values with changes in link utilization on an SDN testbed. Each sample statistics is taken over a duration of 30s.

First we compare our proposed architecture with random and cyclic random OCN assignments as described in [2]. These are used in comparing the performance of our strategy. In the random scenario hosts are assigned either optical or packet depending on value random number obtained from the seed. Traffic to destination is sent randomly in a LogNormal distribution model. We then implement a random cyclic traffic with stability time of 16s. Each flow is assigned to a EPN or OCN after the period the destination host is changed. Figure 8 shows our technique of Laplace spectral distance outperformed the two random assignments described in average throughput. This we attributed to the local traffic connectivity effects as traffic from other flows sharing common incidence source vertex are also considered. We also investigated the effect of including flow connectivity in selecting a flow on average configuration delay. The topology in figure 6 is used to measure cumulative delay using Cloudsim simulator. As shown in figure 9 our proposed flow selection technique considering logical connectivity shows a better performance over the selection without considering logical connectivity.

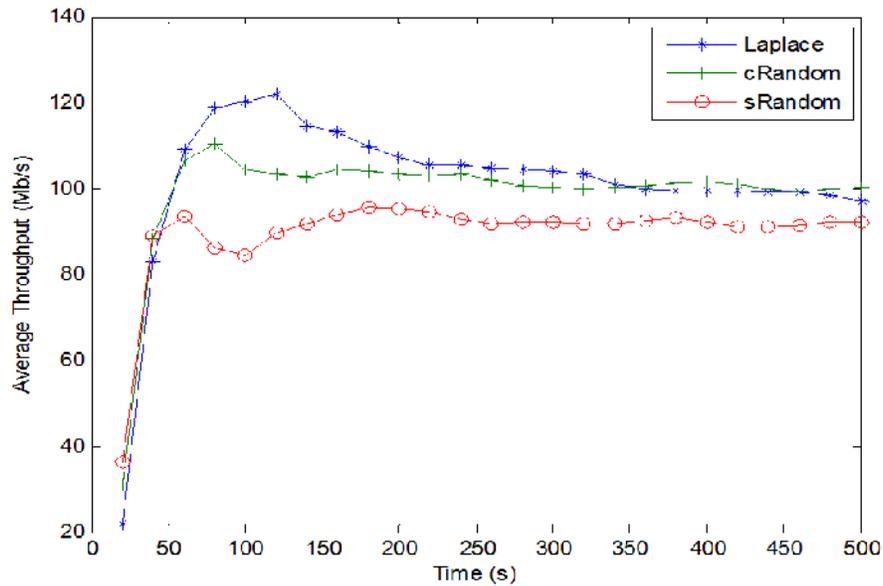

Figure 8. Result showing the effect of flow selection techniques on throughput. Four different strategies are compared. LSD is Laplace spectral distance, Cyclic random with permutation value 6, and seeded random.

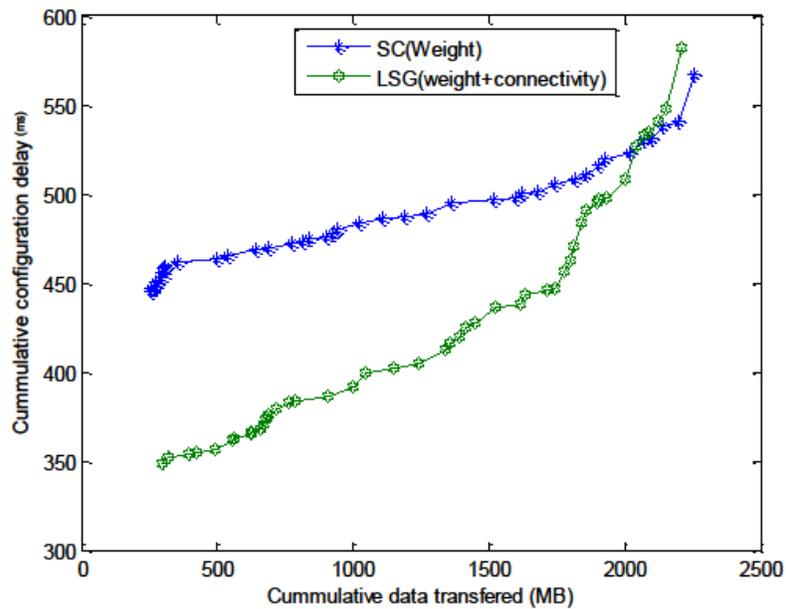

Figure 9. Comparing the effectiveness of adopting a connectivity traffic measure with simple traffic difference. Configuration delay is measured in Milli seconds (ms)

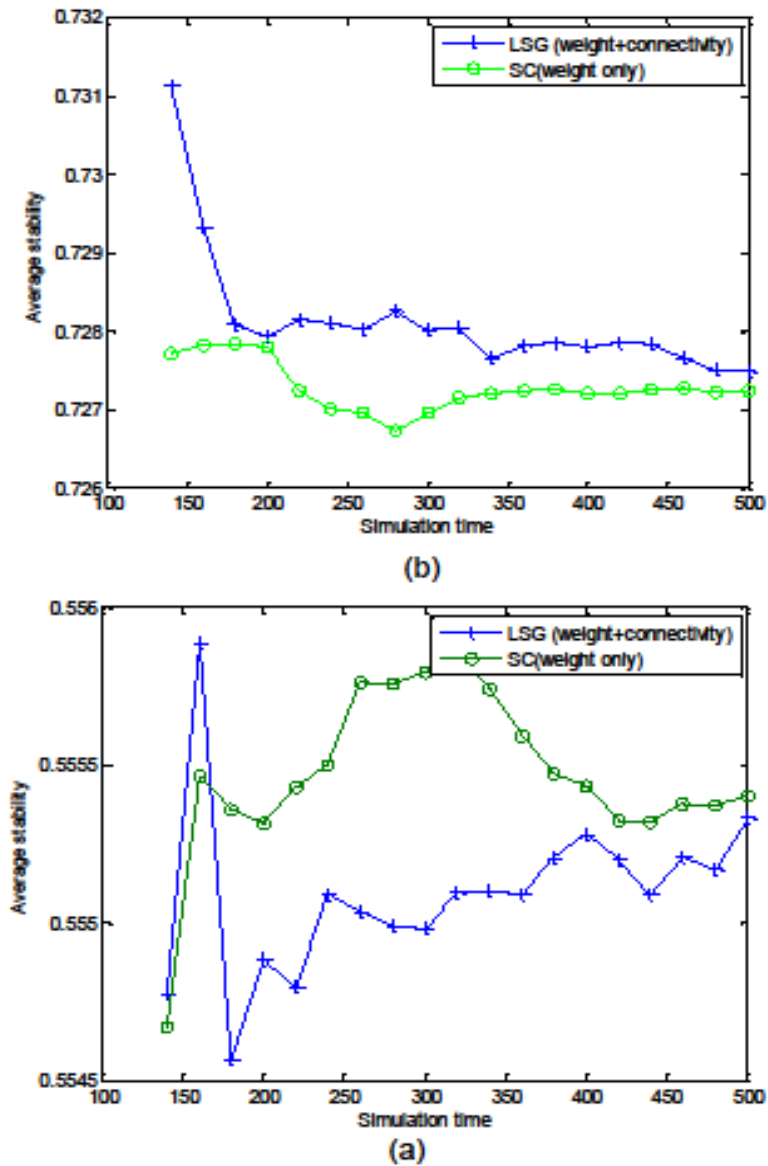

Figure 9. Running the simulation experiment with 4 and 16 wavelengths. Stability increases with wavelength size

Having established the improvements on random techniques by our scheduling model, we then turn to comparing the effect of wavelength size on the performance of the strategy. We define the stability of an optical path as the coefficient that indicate route reused by a flow. Higher flow stability means consistent use of a route by a flow and lower route tear down. As wavelength size increases, the stability of our proposed strategy also increases. Figure 10 shows the result obtained from running a simulation with 4 and 16 wavelengths. At various cases the stability increases with wavelength size.

## 6. CONCLUSIONS

In this paper we proposed a traffic management strategy suitable for electrical packet and circuit optical. We present technique for detection and selection of traffic flows for offload from electrical into packet networks. We also formulate a ranking mechanism for candidate flows to be moved into optical network. Our work also demonstrated how flows identification and selection improved two important parameters for hybrid network - stability and throughput. Our work complement the growing research interest in hybrid electrical packet and circuit networking by presenting traffic change detection, flows selection, and network resource allocation strategies.

## AUTHORS

Ibrahim K. Musa: Received his B.Sc. (Hons.) degree in computer science and masters from Federal University of Technology Nigeria in 2006 and 2009 respectively. He was employed as lecturer in the Federal University of Technology Nigeria in 2008 and worked as a Cisco and Mic rosoft instructor in 2007-2009. In 2010 he started his Ph.D. in Computer Science and Electronic Engineering department of University of Essex, United Kingdom. His current research interest is Resource virtualization in cloud Computing

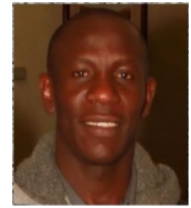

Stuart D. Walker received the B.Sc. (Hons.) degree in physics from Manchester University, Manchester, U.K., in 1973 and the M.Sc. and Ph.D. degrees in electronic systems engineering from the University of Essex, Essex, U.K., in 1975 and 1981, respectively. After completing a period of contractual work for British Telecom Laboratories between 1979 and 1982, he joined the company as a Staff Member in 1982. He worked on various aspects of optical system design and was promoted to Head of the Regenerator Design Group in 1987. In 1988, he became Senior Lecturer and then Reader in Optical Communication at the University of Essex. He has led eight patents and authored over 140 publications. In 2004, he was promoted to a Professorship and is currently Head of the Optical Systems Laboratory at the university.
His current interests focus on modelling and analysis of advanced optical network components and systems.

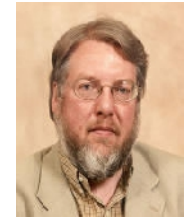